\documentclass[tradiabstract,letter]{aa} 
\usepackage{natbib} 
\usepackage{graphicx}
\usepackage{amssymb}
\usepackage{amsmath}
\usepackage{txfonts}
\usepackage{epsfig}
\usepackage{enumitem}

\def\swift{{\em Swift}}

\begin{document}

   \title{On the super-orbital modulation of supergiant high mass X-ray binaries}

   \author{E. Bozzo 
          \inst{1} 
          \and L. Oskinova
          \inst{2,3}     
          \and A. Lobel 
          \inst{4}
          \and W.-R. Hamann
          \inst{2}                        
          }
   \institute{Department of Astronomy, University of Geneva, Chemin d’Ecogia 16,
             CH-1290 Versoix, Switzerland; \email{enrico.bozzo@unige.ch}    
          \and 
         Institut f\"ur Physik und Astronomie, Universit\"at Potsdam, Karl-Liebknecht-Strasse 24/25, 14476 Potsdam, Germany
          \and 
          Kazan Federal University, Kremlevskaya Str., 18, Kazan, Russia 
          \and 
          Royal Observatory of Belgium, Ringlaan 3, 1180 Brussels, Belgium.
             }
   
   \date{Submitted: -; Accepted -}

  \abstract{The long-term X-ray lightcurves of classical supergiant X-ray binaries and supergiant fast X-ray transients show relatively 
  similar super-orbital modulations, which are still lacking a sound interpretation. We propose that these modulations are related to 
  the presence of corotating interaction regions (CIRs) known to thread the winds of OB supergiants. To test this hypothesis, we couple 
  the outcomes of 3-D hydrodynamic models for the formation of CIRs in stellar winds with a simplified recipe for the accretion onto 
  a neutron star. The results show that the synthetic X-ray light curves are indeed modulated by the presence of the CIRs. The exact 
  period and amplitude of these modulations depend on a number of parameters governing the hydrodynamic wind models and on the binary 
  orbital configuration. To compare our model predictions with the observations, we apply the 3-D wind structure previously shown to 
  well explain the appearance of discrete absorption components in the UV time series of a prototypical B0.5I-type supergiant. Using 
  the orbital parameters of IGRJ16493-4348 which has the same B0.5I donor spectral type, the period and modulations in the 
  simulated X-ray light-curve are similar to the observed ones, thus providing support to our scenario. We propose, that the 
  presence of CIRs in donor star winds should be considered in future theoretical and simulation efforts of wind-fed X-ray binaries.}


   \maketitle

\section{Introduction}
\label{sec:intro}

Supergiant high mass X-ray binaries (SgXBs) are a sub-class of high mass X-ray binaries (HMXBs) hosting 
a compact object and an OB supergiant star \citep[see][for a recent review]{walter15}. 
SgXBs are typically divided in two groups, i.e., the classical systems, which show a virtually persistent high X-ray luminosity, and the 
supergiant fast X-ray transients (SFXTs), which feature a still highly debated and peculiarly prominent variability in X-rays 
\citep{nunez17}. The bulk of the X-ray radiation 
from the SgXBs can be reasonably well explained as being due to the accretion of the stellar wind from the OB supergiant 
onto a highly magnetized neutron star (NS), with no evidence supporting the presence of long-lived accretion disks \citep{bozzo08,shakura12,romano15,hu17}. 
No clear indication has yet been reported of systematic differences between the 
properties of the supergiant stellar winds in classical systems and SFXTs (Hainich et al. 2017, in preparation), 
which also share similar orbital periods \citep{lutovinov13,bozzo15} 
and super-orbital modulations \citep[see Table~\ref{tab:tab1} and][]{corbet13}.   
 \begin{table}[t!]
   \scriptsize	
 \begin{center} 	
 \caption{SgXBs with confirmed  
 orbital, $P_{\rm NS}$, and super-orbital, $P_{\rm SO}$, periods from \citet{corbet13}.} 	
 \label{tab:tab1} 	
 \begin{tabular}{@{}llll@{}} 
 \hline 
 \hline 
 \noalign{\smallskip}  
 Source Name & class & $P_{\rm NS}$ & $P_{\rm SO}$  \\ 
            &  & (days) & (days)      \\
 \noalign{\smallskip} 
 \hline 
 \noalign{\smallskip} 
IGR\,J16479-4514	& SFXT & 3.3199$\pm$0.0005 & 11.880$\pm$0.002 \\
\noalign{\smallskip}
IGR\,J16418-4532	& SgXB & 3.7389$\pm$0.0001 & 14.730$\pm$0.006 \\
\noalign{\smallskip}
4U\,1909+07	& SFXT & 4.4003$\pm$0.0004 & 15.180$\pm$0.003 \\
\noalign{\smallskip}
IGR\,J16493-4348	& SgXB & 6.782$\pm$0.001 & 20.07$\pm$0.01 \\
\noalign{\smallskip}
2S\,0114+650 & SgXB & 11.591$\pm$0.003 & 30.76$\pm$0.03 \\
\noalign{\smallskip}
  \hline
  \end{tabular}
  \end{center}
  \end{table}
  
The super-orbital variability in disk accreting X-ray binaries is usually ascribed to the precession of the disk or to the precession of the compact object 
in its center \citep[see, e.g., the cases of Her\,X-1, SMC\,X-1, and LMC\,X4;][]{petterson75,ogilvie01,pot13}, but a solid interpretation of the same  
phenomenon in wind-fed binaries is still lacking. The lightcurves of the latter folded on their 
super-orbital period display a large variety of morphologies and modulations that are stable over years, although the sources 
are generally thought to be accreting from a much less regularly structured environment compared to that provided by an accretion disk. 
In the case of the classical SgXB 2S\,0114+650, \citet{farrell08} 
reported on the detection of spectral slope changes as a function of the super-orbital phase but could not detect corresponding variations in the absorption column 
density \citep[also due to the limited coverage at energies $\lesssim$3~keV provided by the instruments on-board RXTE;][]{bradt93}. 
These authors suggested that the most likely cause of the super-orbital variability was a modulation of the mass loss rate from the supergiant star. 
For all other systems showing a super-orbital modulation, the only X-ray data providing coverage on different super-orbital phases
are those collected with \swift/BAT \citep{gehrels05, barthelmy05}. The relatively low signal-to-noise ratio of these data and the 
energy band-pass limited to $\gtrsim$15~keV hampered so far any investigation of spectral variability as a function of the super-orbital phase. 
\citet{koeni06} proposed that the mass accretion rate modulation could be produced as a consequence of tidal interaction-driven 
oscillations of the supergiant star, but the authors showed that such mechanism only works for strictly circular orbits.  
\citet{corbet13} also discussed the possibility that the super-orbital modulations are due to a third body orbiting 
the inner massive binary. However, the same authors highlighted that a stable three body solution requires a hierarchical system with the third body 
in a very distant orbit, while all super-orbital modulations discovered so far in SgXBs are not longer than roughly 3 times the orbital periods of these sources.  
  
We propose here that the super-orbital periodicities in classical SgXBs and SFXTs are produced as 
a consequence of the interaction between the compact object with the so-called ``corotating interaction regions'' 
threading the winds of OB supergiants.

\section{Corotating interaction regions around OB supergiants}
\label{sec:cir}

The stellar winds of OB supergiants are well known to be characterized 
by complex velocity and density structures \citep[see, e.g.,]
[and references therein]{puls08}. The smaller structures, i.e. ``clumps'', 
are typically endowed with an increased density of a factor of $\gtrsim$10 compared to a  
smooth wind and can be as large as $\sim$0.1~$R_*$, where $R_*$ is 
the OB supergiant radius. Clumps are usually invoked to interpret the stochastic X-ray variability displayed 
by SgXBs on time scales of 10-1000~s \citep[see, e.g.,][and references therein]{nunez17}. 

The existence of larger structures in the OB supergiant winds was 
suggested in the early 80s \citep{mullan84}, and confirmed by the detection of discrete absorption 
components \citep[DACs; see, e.g.,][]{new}. These features are observed to propagate blue-ward 
on time-scales comparable with the stellar rotation through the profiles of  
UV resonance lines in OB supergiants \citep{massa95,prinja98}. \citet{cranmer96} used hydrodynamic models to show that 
irregularities on the stellar surface related either to dark/bright spots, 
magnetic loops, or non-radial pulsations can lead to the formation of 
corotating interaction regions (CIRs) causing 
spiral-shaped density and velocity perturbations in the stellar wind up to several 
tens of stellar radii. The CIRs are invoked to explain modulations of
the X-ray emission observed in single OB stars
\citep{Oskinova2001, Naze2013, Massa2014}. 

An advanced model to reproduce the observational properties of DACs in OB supergiants with CIRs was developed by 
\citet{lobel08}, who also investigated the dependence of the extension, velocity, and density profile 
variations of these structures as a function of the different properties of the 
stellar surface spots from which they originate. 
The intensity of the spot, its size, and its rotational velocity (that could in principle be different from that of the star) are the main 
parameters regulating the density/velocity contrast of the CIR compared to the surrounding unperturbed stellar wind, typically limited to a factor of a few. 
A larger rotational velocity of the spot also increases the winding of the spiral arms. 
The available observations of the UV variability of massive star resonance lines can be used to constrain all the spot free 
parameters of the model.  

\begin{figure}
\centering
\includegraphics[scale=0.032]{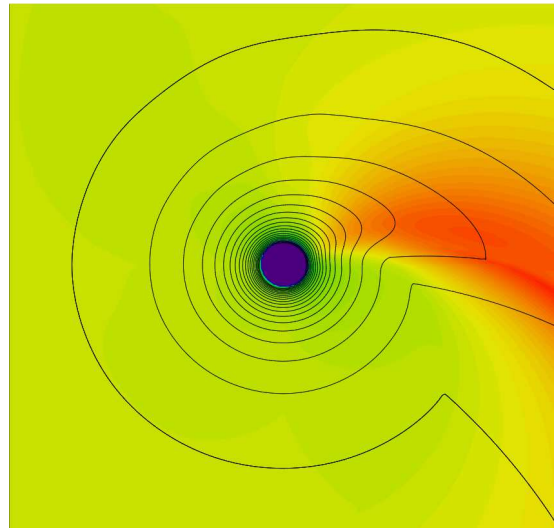}
\caption{An example of hydrodynamic calculations of the CIR in the wind of  
the B0.5 Ib supergiant J\,Pup. There is a bright spot on the stellar surface producing  
a single CIR. The spot is assumed to be 
20\% brighter than the stellar surface, and is characterized by an angular diameter  
of 20$^{\circ}$. A full revolution of the CIR occur in $\sim$10.3~days. 
The colors code the wind density of the CIR model, relative to the density of the unperturbed, smooth wind. The maximum over-density 
in the CIR is by a factor 1.22 (red colors). In wind regions shown in yellow this ratio hardly deviates from unity, while in the green 
regions the wind is rarefied by a factor below 0.97. 
The black solid drawn lines show 20 overplotted contours of equal radial velocity in the hydrodynamic 
rotating wind model. The outermost curve marks the isovelocity line at an outflow velocity of 
1460~km~s$^{-1}$, while the curves at smaller distances from the stellar surface are shown for decreasing 
steps of 73~km~s$^{-1}$. The case with two CIRs mentioned later in this paper is produced by assuming 
an additional spot that is 8\% brighter than the surface of the supergiant star and has an angular diameter of 30$^{\circ}$.}
\label{fig:fig1} 
\end{figure}

We show in Fig.~\ref{fig:fig1} a hydrodynamic  simulation of \citet{lobel08} obtained
from the application of the Zeus-3D code for a radiatively-driven rotating wind in the B0.5~Ib supergiant J\,Pup (HD\,64760). 
This star is characterized by a mass of 20~$M_{\odot}$, a radius of 22~$R_{\odot}$, a mass loss 
rate of 9$\times$10$^{-7}$~$M_{\odot}$~yr$^{-1}$, and a terminal wind velocity of 1500~km~s$^{-1}$ \citep[see Table~1 of][]{lobel08}. 

\section{The CIR induced super-orbital modulation in SgXBs}
\label{sec:so}

In order to show how the presence of CIRs around the supergiant star can introduce a super-orbital modulation 
of the X-ray luminosity from a SgXB, we assume here a simplified NS accretion scenario, following the treatment in \citet{oskinova12}. 
The cross section of the NS for the capture of the stellar wind material is provided by the so-called 
accretion radius, $r_{\rm accr}=\frac{2Gm_{\rm X}}{\varv_{\rm rel}^2}$, 
where $m_{\rm X}$ is the NS mass and $\varv_{\rm rel}$ is the relative velocity 
between the NS and the massive star wind. For the CIR, both the radial and tangential 
components of the velocity at the NS location are taken into account in the computation.  
The mass accretion rate onto the NS is given by $S_{\rm accr}=4\pi \zeta \frac{(Gm_{\rm X})^2}{\varv^3_{\rm rel}}\rho$,  
and the correspondingly released X-ray luminosity is $L^{\rm d}_{\rm X}=\eta S_{\rm accr} c^2$.
In the equations above, we indicated with $c$ the speed of light, with $\rho$ the local wind density (derived from the outputs of the 
hydrodynamic model), and considered for all cases of interest $\zeta\sim 1$ and $\eta\sim 0.1$ \citep[$\zeta$ is a parameter 
included to take into account corrections related to the contribution of the radiation pressure and the finite cooling time of the gas, 
while $\eta$ parametrizes the efficiency of accretion onto a NS; see, e.g.,][]{ostriker}. It is clear from this simplified treatment 
that both the density and velocity contrasts of the CIR compared to the smooth wind can affect the resulting X-ray luminosity, as they can 
significantly alter the mass accretion rate and the size of the NS cross section for the capture of the wind material. 

Let's consider first a case in which the NS orbital period is given by $P_{\rm NS}$ and a single CIR rotating 
with a period $P_{\rm CIR}$ is present in the wind of the supergiant companion. A difference between $P_{\rm NS}$ and $P_{\rm CIR}$ can be expected in a  
not-synchronously rotating binary \citep[see, e.g.,][and references therein]{koeni06}, or in case the stellar spot does not rotate with the same 
velocity of the supergiant star \citep{lobel08}. The mass accretion rate onto the NS is altered every time the compact object encounters  
the CIR along its orbit and the amplitude of the variation is regulated by the CIR velocity/density contrast compared to the rest of the stellar wind. 
The period of the super-orbital modulation, $P_{\rm SO}$, is thus 
given by the difference between the NS and the CIR angular velocities: 
\begin{equation}
 P_{\rm SO}=\frac{1}{\left|\frac{1}{P_{\rm NS}}-\frac{1}{P_{\rm CIR}}\right|}=
 \left|\frac{P_{\rm NS} P_{\rm CIR}}{P_{\rm CIR}-P_{\rm NS}}\right|.
 \label{eq:po} 
\end{equation}

\begin{figure}
\centering
\includegraphics[scale=0.57]{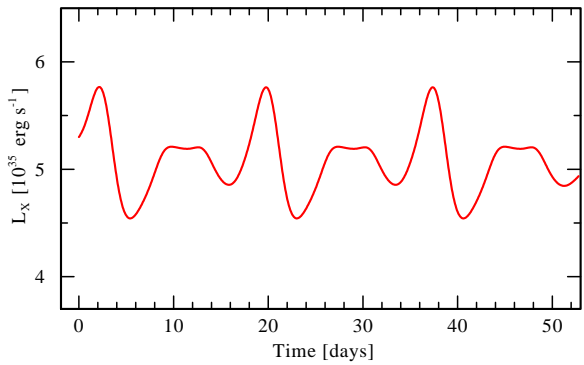}
\caption{Simulated long term lightcurve of IGR\,J16493-4348, assuming a single CIR in this system rotating 
with a period of $\sim$10.3~days.} 
\label{fig:fig2} 
\end{figure}

As an example, we consider in detail the case of the classical SgXB IGR\,J16493-4348. The donor star in this system has the same  
spectral type as J\,Pup and the estimated distance is 6-26~kpc \citep[see][and references therein]{nunez17}. 
Given the typical parameters of a B0.5 Ib star (Sect.~\ref{sec:cir}) and the measured orbital period 
of the NS in IGR\,J16493-4348 (Table~\ref{tab:tab1}), the separation between the compact object and the supergiant is $\sim$1.8~$R_*$. 
According to Eq.~\ref{eq:po}, a single CIR rotating with a period of $\sim$10.3~days is thus expected to give rise to a super-orbital period 
of $\sim$20~days, as indeed observed in this source. The X-ray lightcurve simulated using the output of the hydrodynamic 
model for the CIR in J\,Pup (Fig.~\ref{fig:fig2}) is characterized by an average luminosity  
which is in good agreement with the observations of IGR\,J16493-4348\footnote{The absolute value 
of the average X-ray luminosity from IGR\,J16493-4348 ranges from 
$\sim$10$^{35}$~erg~s$^{-1}$ to $\sim$10$^{36}$~erg~s$^{-1}$, depending on the poorly known distance to the source (6-26~kpc).}.  
The amplitude of the modulation is smaller than that observed from IGR\,J16493-4348, but the exact value depends on, e.g., the brightness of 
the spot on the stellar surface responsible for the CIR generation (in the present case the hydrodynamical simulations were tuned 
to reproduce the results UV spectroscopic monitoring for J Pup). 

\begin{figure}
\centering
\includegraphics[scale=0.57]{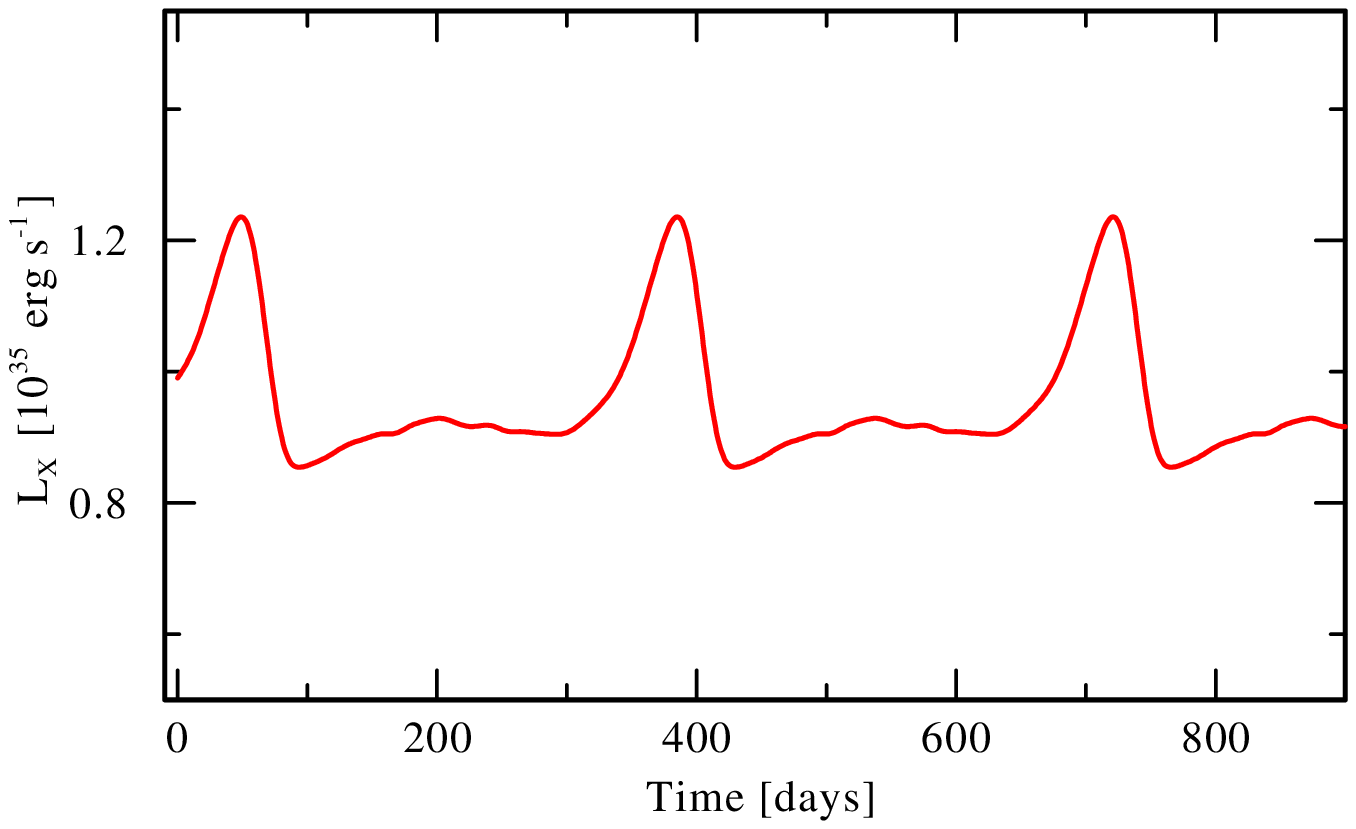}
\includegraphics[scale=0.57]{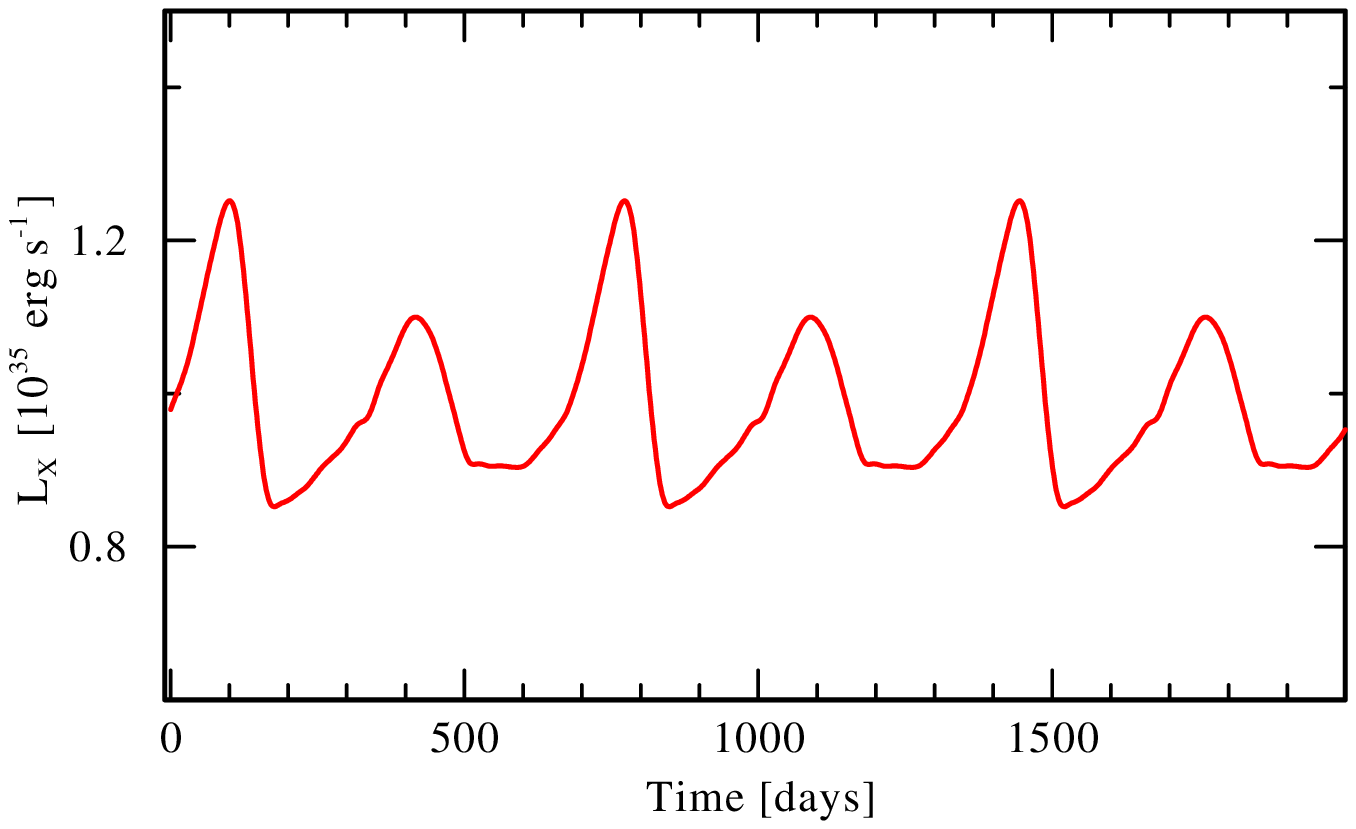}
\caption{{\it Top}: Same as Fig.~\ref{fig:fig2} but assuming an orbital separation of 2.5~$R_*$, corresponding to an orbital period of $\sim$10.6~days. As this is 
only slightly larger than the CIR period ($\sim$10.3~days), the super-orbital modulation occurs on a much longer time-scale ($\sim$330~days).  
{\it Bottom}: Same as above but using 2 CIRs. The super-orbital modulation is now characterized by a double asymmetric 
peak repeating every $\sim$660~days, corresponding to the fact that the NS is crossing two CIRs with different density/velocity contrasts 
along its orbit.}
\label{fig:fig3} 
\end{figure}

In a situation in which multiple CIRs are present and intercept the plane of the NS orbit, it would be naturally expected that their density/velocity contrasts  
are significantly different, reflecting the distinct properties of the stellar surface spots from which they originated. 
Assuming that $n$ CIRs cross the NS orbit, the overall period of the super-orbital modulation will be in this case $n$$\times$$P_{\rm SO}$. 
Several peaks of variable intensities can thus be originated in the modulation profile depending on the different CIR density/velocity contrast  
at the NS location. We show in the bottom plot of Fig.~\ref{fig:fig3} an example in which a second CIR is included in our 
calculations, giving rise to a double-peaked super-orbital modulation (see also the caption of Fig.~\ref{fig:fig1}). 
A multiple-peaked super-orbital modulation seems particularly interesting to reproduce structured profiles of the folded X-ray lightcurves 
displayed by, e.g., the classical SgXB 4U\,1909+07 and the SFXT IGR\,J16479-4514 \citep[see Fig.~8 and 10 of][]{corbet13}.
Figure~\ref{fig:fig3} shows that the super-orbital 
variability is critically depending on the difference between $P_{\rm NS}$ and $P_{\rm CIR}$. Although CIRs seem to be an ubiquitous 
property of all supergiant stars \citep[see, e.g.,][and references therein]{massa15}, the detectability of a super-orbital modulation in the 
currently proposed model could be hampered in all those unfavorable cases where $P_{\rm NS}\simeq P_{\rm CIR}$ and $P_{\rm SO}$ exceeds  
the available observational time-span for a SgXB.  

Although all the free parameters on the number and properties of the CIRs can be fine tuned to obtain a reasonable match with the 
properties of the super-orbital modulations of all sources in Table~\ref{tab:tab1}, there is not an obvious way to explain the empirical relation 
connecting the system orbital and super-orbital period as shown by \citet[see their Fig.~1;][]{corbet13}. If the relation will be confirmed by future observations, 
a better understanding of the CIR formation in the SgXBs is required to investigate the possibility of explaining this observational finding 
in the current model. 

We neglected in the present simplified approach the role of the NS spin period and magnetic field, as well as the effect of 
eccentric orbits and the interaction between the X-rays from the NS and the stellar wind. 
A strong NS magnetic field and slow spin period can induce large modulations of the X-ray luminosity, 
due to the onset of magnetic and/or centrifugal gates \citep{grebenev07,bozzo08}. While this is relevant to explain the short 
time-scale X-ray variability ($\sim$10-1000~s) displayed by classical SgXBs and SFXTs \citep{bozzo16}, we expect this variability to be 
largely smoothed out when considering the much longer integration times corresponding to the super-orbital modulations (tens of days). 

As described in \citet{bozzo16}, the lack of a proper treatment of the X-ray illumination of the stellar wind by the accreting compact 
object limits the validity of the outcomes of the calculations to low luminosity SgXBs ($L_{\rm X}$$\lesssim$10$^{35}$~erg~s$^{-1}$) and can 
only provide indications in case of brighter systems.  This effect is thus unlikely to be critical in the case of the SFXTs, as their average X-ray luminosity 
is generally far below the critical level required to produce a systematic disruption of the stellar wind on scales that are as large as those expected for the CIRs 
\citep[see, e.g.,][and references therein]{ducci10}. However, this might not apply to bright persistent classical wind-fed SgXBs with average X-ray luminosities 
$\gg$10$^{36}$-10$^{37}$~erg~s$^{-1}$ \citep[e.g., Vela\,X-1;][]{watanabe06,sander17}, which we predict to not display super-orbital modulations. 
Note that all SgXBs discovered so far to display a super-orbital modulation are characterized by relatively low long-term luminosities, the brightest being 
IGR\,J16493-4348 with an estimated average X-ray luminosity of $\sim$1.5$\times$10$^{36}$~erg~s$^{-1}$ when the largest allowed distance of 26~kpc 
is considered. 

Finally, the usage of circular orbits in our calculation 
was adopted to limit the number of required hydrodynamic runs and to provide more intuitive examples to promote the proposed scenario. 
However, we do not expect that our conclusions will change significantly with the introduction of a relatively small eccentricity 
\citep[$\lesssim$0.2;][and references therein]{corbet13}, 
as the range of orbital separations spanned by the NSs would be limited and the density/velocity contrasts within each CIRs 
are expected to undergo major variations only on several stellar radii (for reasonable assumptions of the model parameters).

\section{Discussion and conclusions}
\label{sec:conclusion}

We proposed in this letter that the still poorly understood super-orbital variability displayed by several classical SgXBs and SFXTs 
is related to the presence of CIRs in the winds of their OB supergiants. The mechanisms leading to the formation of CIRs are not 
yet fully understood and the model considered in Sect.~\ref{sec:cir} exploits the most widely accepted idea that CIRs are driven by spots  
on the supergiant surface with a number of free parameters that can be adjusted to reproduce the observational 
properties of DACs in many isolated OB supergiants. At present, no observations of DACs are available for the donor stars 
in classical SgXBs and SFXTs, and it is thus difficult to present an exhaustive exploration of the entire model parameter space 
(this is beyond the scope of this work, given also the computationally expensive runs of the 
hydrodynamic simulations). However, the examples provided in Sect.~\ref{sec:so} show that the number and physical characteristics of the CIRs can be adjusted  
within reasonable boundaries to reproduce the main observed properties of the super-orbital modulations displayed by all sources in Table~\ref{tab:tab1}. 
This paves the way to future theoretical and simulation efforts exploring the proposed model in more details and overcoming the  
simplifications adopted in the current approach, including the presence of the X-ray irradiation of the stellar wind and eccentric orbits.   

We remark that an important open question in the context of the present interpretation of the super-orbital modulations is if CIRs can be stable for years. 
Indeed, most of the DAC observations leading to the concept of CIRs were carried out as part of the MEGA campaign \citep{mega} performed with the 
IUE satellite \citep{iue} and no longer repeated afterwards. Long dedicated monitoring campaigns of the UV spectroscopic variability 
of the previously observed supergiant stars and the other supergiants hosted in the SgXBs are thus critically required.

\section*{Acknowledgments}

We thank the anonymous referee for constructive comments which helped to improve the paper. 
This publication was motivated by a team 
sponsored by the ISSI in Bern,
Switzerland. EB and LO thank ISSI for the financial
support during their stay in Bern. EB is grateful for the 
hospitality of the Institut f\"ur Physik und Astronomie 
(Universit\"at Potsdam) during part of this work. EB 
acknowledges financial traveling contribution from 
the Swiss Society for Astronomy and Astrophysics.  
LO acknowledges support by the DLR grant 50\,OR\,1302 and partial
support by the Russian Government Program of Competitive 
Growth of Kazan Federal University. AL acknowledges partial financial support by the Belgian Federal 
Science policy Office under contract No. BR/143/A2/BRASS and by the ESA-Gaia Prodex Programme 2015-2017. 

\bibliographystyle{aa}
\bibliography{superorbital}

\end{document}